\documentclass{ifacconf}

\usepackage{graphicx}      % include this line if your document contains figures
\usepackage{amsmath}
\usepackage{mathtools} 
\usepackage{amsfonts}
\usepackage{amssymb}
\usepackage[hyphens]{url}

\usepackage{natbib}        % required for bibliography
\usepackage{siunitx} 
\usepackage{xspace}
\usepackage{xcolor} 
\usepackage{xfrac}
 
\newcounter{example}
\newenvironment{example}[1][]{\refstepcounter{example}\par\vspace{1mm}
	\noindent \textbf{Example~\theexample. #1} \rmfamily}{}

\newcommand{\eg}{e.g.,\xspace}
\newcommand{\ie}{i.e.,\xspace}
\newcommand{\cf}{cf\/.\/~}
\newcommand{\iid}{i.i.d.\xspace}

\newcommand\figref[1]{Fig.~\ref{#1}}

\newcommand\secref[1]{Sec.~\ref{#1}}
\usepackage{ifthen}
\newboolean{authnotes}
 \setboolean{authnotes}{false}

\ifthenelse{\boolean{authnotes}}
{
\newcommand{\jm}[1]{\footnote{{\bf\color{red} Jose: #1}}}
\newcommand{\db}[1]{\footnote{{\bf\color{blue} Dominik: #1}}} 
\newcommand{\st}[1]{\footnote{{\bf\color{green!80!black} Sebastian: #1}}}
}
{
\newcommand{\jm}[1]{}
\newcommand{\db}[1]{}
\newcommand{\st}[1]{}
}

\usepackage{fancyhdr}
\newcommand{\mytitle}{\textbf{Accepted final version.}
	To appear in \textit{Proc. of the 8th IFAC Workshop on Distributed Estimation and Control in Networked Systems, 2019}.\\
	\copyright 2019, IFAC.}
\begin{document}

\begin{frontmatter}

\title{Predictive Triggering for Distributed Control \\of Resource Constrained Multi-agent Systems}

\thanks[footnoteinfo]{This work was supported in part by the German Research Foundation (DFG) within the priority program SPP 1914 (grant TR 1433/1-1), the Cyber Valley Initiative, and the Max Planck Society.}

\author[First]{Jos\'{e} Mario Mastrangelo}
\author[First,Second]{Dominik Baumann} 
\author[First]{Sebastian Trimpe}

\address[First]{Intelligent Control Systems Group, Max Planck Institute for Intelligent Systems, Stuttgart, Germany. (e-mail: $\lbrace$mastrangelo, trimpe$\rbrace$@is.mpg.de, dbaumann@tuebingen.mpg.de)}
\address[Second]{Corresponding Author}

\begin{abstract}    % 50-100 words.
A predictive triggering (PT) framework for the distributed control of resource constrained multi-agent systems is proposed. 
By predicting future communication demands and deriving a probabilistic priority measure, the PT framework is able to allocate limited communication resources in advance.
%The PT framework derives a probabilistic priority measure, according to the probability of by predicting future communication demands.
%\textit{Under resource constraints, event-triggered designs are unable to reallocate...}
%Simulations of a cooperative adaptive cruise control system and experiments on multi-agent cart-pole systems demonstrate the effectiveness of the PT framework over other event-triggered designs at reducing network utilization, while also improving the control error of the system.
%The effectiveness of the PT framework over other event-triggered designs is demonstrated to result in reduced network utilization, while also improving the control error of the system. Simulations of a cooperative adaptive cruise control system and experiments on multi-agent cart-pole systems demonstrate these results.
The framework is evaluated through simulations of a cooperative adaptive cruise control system and experiments on multi-agent cart-pole systems. The results of these studies show its effectiveness over other event-triggered designs at reducing network utilization, while also improving the control error of the system.
\end{abstract}


\end{frontmatter}
%==========================================================================

%%%%%%%%%%%%%%%%%%%%%%%%%%%%%%%%%%%%%%%%%%%%%
% Added after final submission
%%%%%%%%%%%%%%%%%%%%%%%%%%%%%%%%%%%%%%%%%%%%%
\thispagestyle{fancy}    % final submitted: empty
\pagestyle{empty}

\section{Introduction}
\label{sec:Intro}
\vspace{-0.25em}
Multi-agent systems often require communication between agents to maintain system-level control.
In traditional control design, sufficient communication resources are taken for granted and the control input is applied periodically, typically at high rates.
Such designs become challenging when information needs to be transmitted over a communication network shared amongst multiple agents.
To reduce the number of samples in such networked control systems (NCS), event-triggered (ET) designs have been developed, see, \eg \cite{IntroET,EventControl}.
In ET control, information is transmitted when certain events occur, \eg some error growing too large.
However, these binary decisions are typically taken instantaneously, \ie communication must be possible for \emph{all} agents at \emph{every} time instant.
In this work, we target a setting where the limited bandwidth of the communication network only allows for a subset of agents to communicate at every time instant.
Therefore, communication decisions need to be taken \emph{in advance}, a method referred to as \emph{predictive triggering} (PT), allowing prioritized scheduling.
%Traditional control techniques, which transmit data periodically, present challenges in the design of networked control systems (NCS). 
%Event-triggered (ET) designs as in \cite{EventControl}, that communicate only when necessary, require resources to be preallocated since communication decision are taken instantaneously. 
%When negative communication decisions occur, resources are unable to be reallocated with typical network technology. 
%A multi-agent system, depicted in~\figref{fig:distributed_PT_control}, where agents communicate over a shared, limited network, is considered in this paper. Under this communication constraint, ET designs are unable to preallocate resources to all agents,

\begin{figure}[t]
	\begin{center}
		\includegraphics[width=8.6cm]{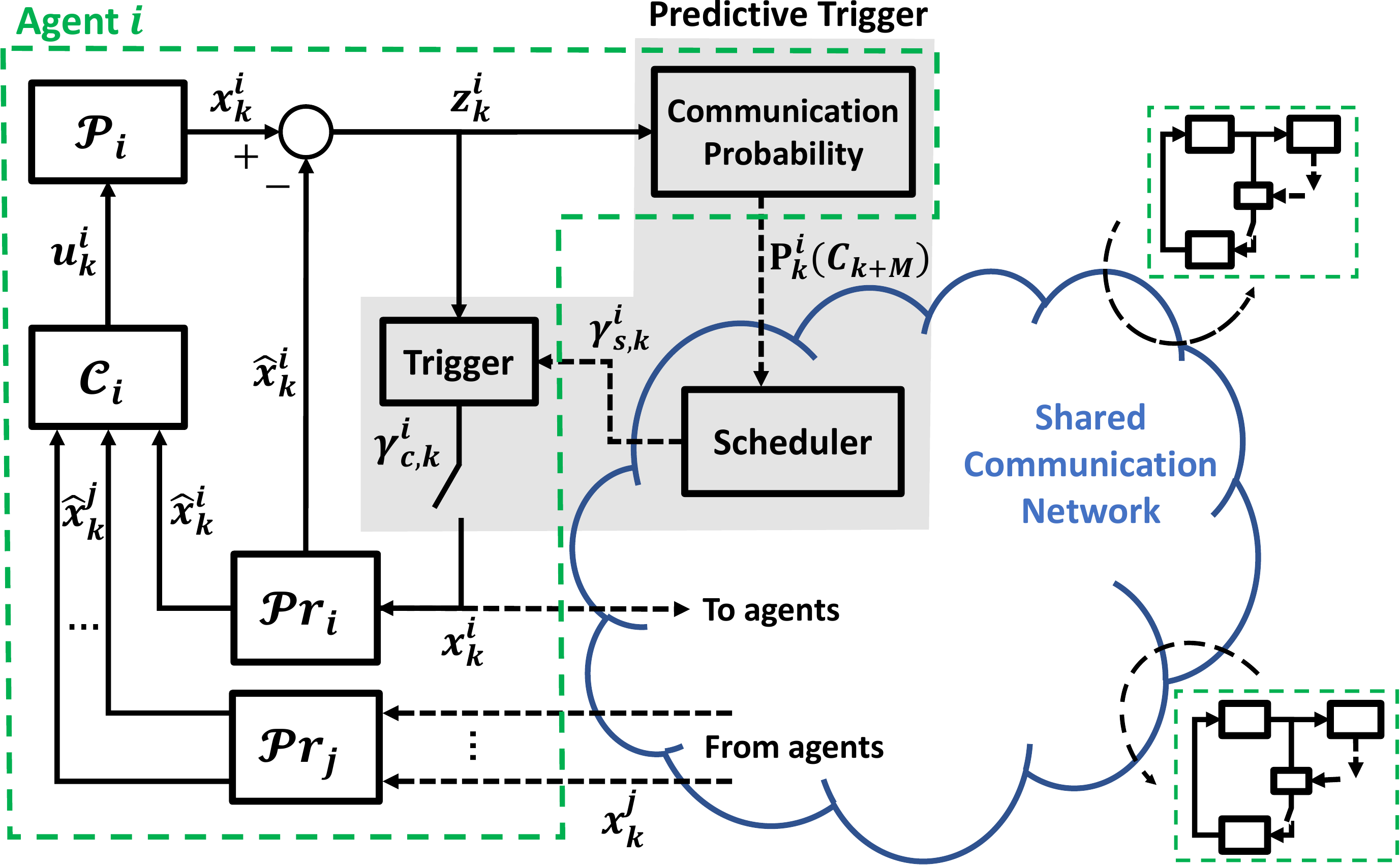}    % The printed column width is 8.4 cm.
		\caption{Multi-agent system connected over a shared network, where each agent $i \in \Omega=\lbrace 1,\ldots,N\rbrace$ is composed of a process $\mathcal{P}_{i}$, predictors $\mathcal{P}r$, and a controller $\mathcal{C}_{i}$. The \emph{Predictive Trigger} (in gray) takes a \emph{Communication Probability} sent by each agent to schedule communication slots to agents (\emph{Scheduler} block). Communication decisions are taken by each agent in the \emph{Trigger} block.}
%		\caption{Multi-agent system connected over a shared network, with each agent $i \in \Omega=\lbrace 1,\ldots,N\rbrace$ composed of a process $\mathcal{P}_{i}$, predictors $\mathcal{P}r_{i}$, $\mathcal{P}r_{j} \ \forall j\in \Omega\backslash\lbrace i\rbrace$, and controller $\mathcal{C}_{i}$. The \emph{Predictive Trigger} proposed herein takes a \emph{Communication Probability} sent by each agent to schedule communication slots to agents (block \emph{Network Scheduler}). Communication decisions are taken by each agent in the \emph{Trigger}.}
		\label{fig:distributed_PT_control}
	\end{center}
\end{figure}

The setup of a distributed multi-agent system communicating over a shared network is depicted in~\figref{fig:distributed_PT_control}.
%Agents attempt to estimate the state of either all, or a subset of system agents, and through communication can receive updates to improve said estimates.
Each agent estimates the state of either all other agents or a subset of the system agents. Through communication, updates are received and help improve said estimates.
By computing a probabilistic priority measure in the \emph{Communication Probability} block, each agent is able to communicate this measure to a centralized \emph{Scheduler}. 
The predictive nature of the setup is such that we compute future communication demands at a horizon, thereby providing the \emph{Scheduler} a time window at which to allocate communication resources, \ie resources are allocated in advance. 
Finally, when allocated a resource, each agent decides whether or not to communicate its state information, in the \emph{Trigger} block, in order to further reduce energy consumption.
%\db{in the end maybe try to save this line. You could for instance write 'each agent estaimtes the state of either all other or a subset of the system agents'. This might already be enough.}
%Depending on the communication topology, agents attempt to estimate the state of either a subset of communicating agents, or all agents, requiring bus-like network protocols such a \cite{LWB}\db{skip reference here. For estimating a bus-like protocol is not necessary. Agents need state of all others or of a subset, estimate it and can receive updates through communication. I wouldn't talk too much about communication topology here, the framework works as long as direct links between the agents that want to communicate exists.}. 
%All agents  attempt to estimate each other's state in order for distributed control to be possible.
\subsubsection{Contributions:} We present the following contributions:
\begin{itemize}
	\setlength\itemsep{0.25mm}
	\item proposal of a PT framework for distributed control of multi-agent systems with limited network bandwidth;
	\item derivation of a probabilistic priority measure for sche-\newline duling limited resources at a prediction horizon; and
	\item comparison of the proposed framework with ET designs, demonstrating improved network utilization and control performance in cooperative adaptive cruise control (CACC) simulations and in multi-agent cart-pole hardware experiments.
\end{itemize}

%CACC code\db{Maybe better code of the simulation example or something like that} and supplementary video\jm{material?}\db{up to you. supplementary material, videos of experiments, all fine with me, can also leave it as it is} available at \url{https://sites.google.com/view/ptformas}.
Simulation code and videos of experiments are available at \url{https://sites.google.com/view/ptformas}.

\subsubsection{Related work:} For general overviews of ET control, we refer the reader to \cite{IntroET,EventControl}.
%while various event generators are proposed in \cite{ChoiceET}\db{I've never cited this paper I think, so this is something you may skip for space reasons.}
In most ET control approaches, communication is triggered instantaneously, with no possibility to reallocate freed resources.
%\ie there is no possibility to reallocate freed resources.
%The development of self- and predictive triggered designs, however, allows for the reallocation of unused resources.
%With the development of self- and predictive triggered designs, communication systems have the ability to reconfigure based on future demands.
%Self-triggering has been proposed to predict future communications in \cite{IntroET}, while PT designs have been proposed for the first time in \cite{PredictiveETSE,ResourceAwareIoT}\jm{remove \cite{PredictiveETSE}}.
Methods of predicting future communication demands have been presented in \cite{IntroET,PredictiveETSE,ResourceAwareIoT} and have largely focused on self-trigger designs, while the latter works also propose the novel concept of predictive triggering for the first time.
These designs, however, still take binary communication decisions requiring that in the worst case all agents can communicate simultaneously.
In emerging applications of NCS, such as formation control of drone swarms, the number of agents wanting to communicate may still exceed the available resources with such designs. This leads to unresolved contention of resource access.
%However, situations may arise\db{This you can word much stronger. Sth like in emerging applications of NCS, such as swarms of drones flying in formation, the number of \ldots} where the number of communicating agents exceeds the available resources, resulting in unresolved contention for access to the limited resources.
Therefore, we build on the general PT but derive a priority measure instead of a binary decision, allowing for prioritized resource allocation.
%\db{In general, this paragraph is nice. But it's not correct to say that self-triggering has been proposed in \cite{PredictiveETSE}, self-triggered control is an older concept that had been proposed already earlier. I would here write sth like predicting future communication demands to allow for rescheduling has been proposed and then continue the way you have it. Then in the end you need to write sth like the general idea of PT (you already introduced this abbreviation, no need to introduce it here again) has been proposed therein for the first time, but taking binary communication decisions}
%\db{here you have to make clear that the binary decision means that in worst case each agent needs to have the possibility to communicate}

%Some of the first works that considered communication decisions in the context of multi-agent control, proposed modeling multi-agent coordination problems using decentralized Markov decision processes~\cite{CommDecCMAS}.  
The inherent resource constraint in large-scale NCS has led to the development of contention resolution algorithms for scheduling limited resources.
%\db{Here it must become clear that this is one approach to deal with the problem we're considering here and what is different in our approach.}
Deterministic scheduling policies have been proposed in \cite{TOD,Molin}, which consider error-dependent prioritization, while a decentralized stochastic policy is considered in \cite{Balaghi}. Decentralized designs address the scalability constraint of NCS but suffer from inevitable collisions.
Herein, we consider a system with a centralized scheduler as in \cite{ErrordependentDS}, where an error-dependent policy is proposed. We, on the other hand, derive a probabilistic priority measure (\ie a purely stochastic policy), and send this probability to the scheduler.
%Initial works by \cite{TOD} introduced the try-one-discard protocol, a deterministic allocation method which prioritized according to estimation error. Randomized scheduling is proposed by \cite{Molin}, while later works proposed using a centralized scheduler (as considered herein) and an error-dependent priority measure for multi-loop NCS in \cite{ErrordependentDS}, and for random access NCS in \cite{Mamduhi}.
%In contrast to these methods, we consider a probabilistic priority measure which sends a probability to the scheduler.
Sending a probability instead of state-information allows us to consider a low dimensional and system independent priority measure, resulting in reduced bandwidth consumption, as we will show in simulation studies and real experiments.
%Using a probability, rather than state information, allows us to consider a low dimensional and system independent priority measure, resulting in a reduction of bandwidth consumption and in simplified scheduling algorithms.
Recently, \cite{Deepcas} proposed DeepCAS, a deep reinforcement learning (RL) algorithm for the control-aware scheduling of NCS. However powerful RL may be, difficulties arise in training and when implementation on embedded hardware is required.

The probabilistic priority measure proposed herein is represented by the probability of communicating at a horizon, and is derived according to the exit times of stochastic processes. 
Exit/stopping time analysis has been applied to arrive at optimal communication decisions in \cite{xu2004optimal,rabi2008optimal}, while \cite{ETPC} and \cite{EventTriggeredLearning} use such analysis to evaluate model performance and trigger learning experiments. 
However, neither design considers predicting future communications with limited resources.
%Using a probability, rather than state information, allows us to consider a low dimensional and system independent priority measure, resulting in a reduction of bandwidth consumption and in simplified scheduling algorithms.

\section{Problem Setting and Main Idea}
\label{sec:ProbSetting}

In this paper, we consider the setup presented in~\figref{fig:distributed_PT_control}, depicting a multi-agent system of $N$ agents communicating over a network with $K\!<\!N$ communication slots, and formulate the resource constrained predictive triggering problem that each agent solves. Herein, a distributed control setting is considered, in which agents communicate their sensor information to other agents over a shared communication network. The components of~\figref{fig:distributed_PT_control} are detailed in the following sections.

\subsection{Process dynamics}
\label{sec:process}
\vspace{-0.6mm}
The uncoupled dynamics of agent $i\in\Omega=\lbrace 1,\dots,N\rbrace$ follow the stochastic, linear time-invariant dynamics
\begin{equation}
\mathcal{P}_{i} \; \mathrm{:} \quad x_{k+1}^{i} = A_{i}x_{k}^{i} + B_{i}u_{k}^{i} + w_{k}^{i},
\label{eq:ProcessControl}
\end{equation}
%with $k\in\mathbb{N}$ the discrete time index, $x_{k}^{i}\in\mathbb{R}^{n_{x}}$ the~state, $u_{k}^{i}\in\mathbb{R}^{n_{u}}$~the input, matrices $A_{i},\,B_{i}$ of appropriate dimension, and initial states $x_{0}^{i}$ chosen from a Gaussian distribution. 
with $k\in\mathbb{N}$ the discrete time index, $x_{k}^{i}\in\mathbb{R}^{n_{x}}$ the~state, $u_{k}^{i}\in\mathbb{R}^{n_{u}}$~the input, and initial states $x_{0}^{i}$ chosen from a Gaussian distribution.
%Noise-free measurements of each agent full state are assumed,
We assume that the full state of each agent can be measured (\ie $y_k^i = x_k^i$), however, state reconstruction via estimation techniques is also possible, as in \cite{solowjowETL19}, but not considered herein.
%\db{I think here it would be important to cite Friedrich's arxiv paper on event-triggered learning. This extension to state estimation is not that trivial}
%and for simplicity, that the state dimension of all agents are the same.
%Further, we assume the dynamics to be disturbed by \iid Gaussian process noise $w_{k}^{i}\!\in\!\mathbb{R}^{n_{x}}$, which captures for instance model uncertainty, and where $w_{k}^{i}\!\sim\!\mathcal{N}(0,\Sigma_{w})$, with constant variance $\Sigma_{w}$.
The process noise $w_{k}^{i}\!\in\!\mathbb{R}^{n_{x}}$, which captures for instance model uncertainty, is assumed to be \iid Gaussian noise, where $w_{k}^{i}\!\sim\!\mathcal{N}(0,\Sigma_{w})$, with constant variance $\Sigma_{w}$.
%\db{is given by may be misleading, this is still our choice to assume such noise. So maybe rather write 'further, we assume the system to be disturbed by Gaussian process nosie' or sth like that}

\subsection{State predictors}
\label{sec:pred}
\vspace{-0.6mm}
At every time step~$k$, each agent may decide to communicate its state to all or a subset of agents if allocated a slot. The binary communication decision~$\gamma_{c,k}^{i}$, denoting positive ($\gamma_{c,k}^{i}\!=\!1$) or negative ($\gamma_{c,k}^{i}\!=\!0$) decisions, determines when agents send their state.
%We assume each agent communicates with a subset of agents and runs predictors $\mathcal{P}r_{j}$
As such, agent~$i$ runs predictors~$\mathcal{P}r_{j}$, to estimate the states of agents $j\in\Omega_{i}\subseteq\Omega\backslash\lbrace i\rbrace$, where $\Omega_{i}$ is the set of agents communicating to agent~$i$. These predictors, initialized with $\hat{x}_{0}^{j}=x_{0}^{j}$, are given by
%\db{This is a bit unclear. We assume each agent communicates with a subset of agents and runs predictors of those agents.}
%These estimates may then be used for control, as will be demonstrated in later sections.
%For the general case\db{eq (2) describes the prediction agent i makes about agent j's state. This is also true for restricted communication patterns, so in my opinion there is no need to assume here that all agents communicate to all others. Just writing sth like agent i's prediction of agent j's state is given by \ldots should be enough} where all agents communicate to all others, \ie $\Omega_{i} = \Omega\backslash\lbrace i\rbrace$, the predictors are given by
%The predictors\db{This I don't understand. $\mathcal{P}r_{j}$ defines the prediction of that agent $i$ makes about agent $j$. So what does predictors mean here? Are there multiple predictors running?} of agents $j$, computed by agent $i$, are given by
%Agent $i$'s prediction of agent $j$'s state is given by
\begin{equation}
\mathcal{P}r_{j} \; \mathrm{:} \quad \hat{x}_{k+1}^{j} =
\begin{cases}
A_{j}\hat{x}_{k}^{j} + B_{j}u_{k}^{j}, & \text{if } \gamma_{c,k}^{j} = 0\\
x_{k+1}^{j},& \text{if } \gamma_{c,k}^{j} = 1,
\end{cases} 
\label{eq:PredictorControl}
\end{equation}
where $\hat{x}_{k}^{j}$ is the predicted state of agent~$j$ at time~$k$~and may be used for control, as will be demonstrated in later sections. In the case when an agent~$j\!\in\!\Omega_{i}$ computes a positive communication decision (\ie $\gamma_{c,k}^{j}\!=\!1$), agent~$i$ resets its prediction of agent~$j$'s state.
%When negative communication decisions occur, the estimate is updated with its previous value, whereas positive communication decisions result in all agents resetting their estimate of agent~$j$'s state.\db{here it's also enough that agent i resets its prediction of agent j in case of communication}
Agent~$i$ can then also run a predictor~$\mathcal{P}r_{i}$ of its own state, similar to~(\ref{eq:PredictorControl}), to determine when deviations occur between its state and prediction.
%\db{ok, seems like for you $\hat{x}_{k}^{j}$ is agent j's prediction of agent i's state. But no matter how you define it, I would still not restrict this discussion to a specific communication pattern. Be careful that your definition is consistent, \ie that you can define all predictors for all agents without ambiguities} 
We define this deviation as the estimation error $z_{k}^{i} \coloneqq x_{k}^{i} - \hat{x}_{k}^{i}$, which resets to zero when communication occurs, and whose Euclidean norm we want to prevent from growing beyond a user-defined threshold $\delta_{i}$. 
%The threshold should be selected according to the process parameters ($A, \Sigma_{w}$), but for the systems considered herein which are either homogeneous or whose state matrices $A$ are similar, we let $\delta_{i}=\delta \ \forall \ i \in \Omega$.
%\begin{equation}
%z_{k}^{i} \coloneqq x_{k}^{i} - \hat{x}_{k}^{i},
%\label{eq:EstError}
%\end{equation}

\subsection{Controller}
\vspace{-0.5mm}
The distributed controller $\mathcal{C}_{i}$ considered herein represents a static linear feedback controller given by
%The general distributed controller $\mathcal{C}_{i}$ that is considered, is given by
%\begin{equation}
%\mathcal{C}_{i} \; \mathrm{:} \quad u_{k}^{i} = f_{i}(\hat{x}_{k}^{i},\hat{x}_{k}^{j}),
%\label{eq:ControlInput}
%\end{equation}
%where $f_{i}$ may be designed as any linear function of the current agents prediction $\hat{x}_{k}^{i}$, as well as the predictions $\hat{x}_{k}^{j} \ \forall \ j\in\Omega_{i}$, that renders the closed-loop system stable.
\begin{equation}
\mathcal{C}_{i} \; \mathrm{:} \quad u_{k}^{i} = F_{i}\hat{x}_{k}^{i} + \sum\nolimits_{j\in\Omega_{i}} F_{j}\hat{x}_{k}^{j}
\label{eq:ControlInput}
\end{equation}
with $\hat{x}_{k}^{i}$ the current agents prediction and $\hat{x}_{k}^{j} \ \forall \ j\in\Omega_{i}$ the predictions of communicating agents, while $F_{i}$ and $F_{j}$ are chosen to render the closed-loop system stable.
%where $f_{i}$ may be designed as any linear function of the system agents' predictions $\hat{x}_{k}^{i}$, for any $i\!\in\!\Omega$, that renders the closed-loop system stable.

\subsection{Network configuration}
\label{sec:NetConfig}

%The communication network considered herein must allow for all agents to communicate information to all others.
%In this paper, we assume the network to be ideal, \ie without delays and packet drops, and to have a limited bandwidth.
In this paper, we assume an ideal network, in that we do not consider delays or packet drops, but have a limited bandwidth.
%\db{A network with limited bandwidth is not ideal. I would write something like we assume the network to be ideal in that we do not consider delays or packet drops, but to have a limited bandwidth.}
Due to the bandwidth limitation, only $K\!<\!N$ communication slots are available for state communication, no matter the trigger design.
A limited bandwidth implies a limited network capacity $\eta$ at each time step (in bytes), which is divided such that $\eta\!=\!4N\!+4n_{x}K$, where $n_{x}$ is the state dimension. This division allows all agents to send scheduling information and $K$ agents to communicate their state at every time step (as 4-byte floats). We can now define the network utilization:
%No matter the trigger design, we assume that a fixed number of communication slots $K$ are available for state communication.
%We assume that no matter the trigger design, that a fixed number of communication slots $K$ are available for state communication. 
%The network capacity is then~divided such that $\eta\!=\!4N\!+4n_{x}K$, where $n_{x}$ is the state dimension and 4-byte floats are used to represent the communicated scheduling and state information. 
\begin{defn}
	Given a network with capacity $\eta$, the network utilization of the system at time~$k$ is defined as
	\begin{equation}
%	U_{k} \coloneqq (bN_{p,k} + 4n_{x}N_{c,k})/\eta.
	U_{k} \coloneqq \frac{bN_{p,k} + 4n_{x}N_{c,k}}{\eta}.
	\label{eq:Utilization}
	\end{equation}
	The size of the scheduling information (in bytes) is denoted by~$b$, while $N_{p,k}\!\leq\!N$, and $N_{c,k}\!\leq\!K$ are the number of agents communicating their scheduling information to the network scheduler and their state, respectively, at time~$k$.
	\label{def:Utilization}
\end{defn}
%Apart from the limited bandwidth, we also assume the network to be ideal, \ie without delays and packet drops.
\vspace{-1em}
\subsection{Main idea}
\label{sec:MainIdea}

Given the limited network, the objective of this paper is to develop a resource-aware scheduling method based on predicted future communication demands. We propose a resource constrained PT framework, which prioritizes resource allocation according to a \emph{probabilistic priority measure}, defined as the probability of wanting to communicate $M$ steps in the future (\ie $M$-step horizon). Each agent computes this probability and communicates it periodically to a \textit{Scheduler}, as in~\figref{fig:distributed_PT_control}, thereby reducing state communication and will be shown empirically to result in net utilization savings.
The \textit{Scheduler} takes allocation decisions $\gamma_{s,k}^{i}$ in advance, according to this priority measure, while the \textit{Trigger} takes state communication decisions. 

\section{Resource Constrained PT Framework}
\label{sec:PTFramework}

In this section, we design a PT framework capable of allocating limited resources in advance, according to a probabilistic priority measure. 
By computing the exit probability of a stochastic process in~\secref{sec:ExitProb}, we are able to derive the probability of communicating at a horizon~$M$ in~\secref{sec:PredHorizon}, and arrive at communication decisions in~\secref{sec:NetMan}.
%The communication probability represents the aforementioned priority measure and is sent to the network scheduler

\subsection{Exit probability}
\label{sec:ExitProb}

%First, we transform~(\ref{eq:ProcessControl}) into a continuous time Ornstein-Uhlenbeck (OU) process $X(t)$. In its integral form, $X(t)$ is expressed as
%In order to obtain tractable exit times, we transform~(\ref{eq:ProcessControl}) into a continuous time Ornstein-Uhlenbeck (OU) process $\lbrace X(t)\colon t\geq0\rbrace$, expressed in its integral form as
%In order to obtain tractable exit times, we consider a continuous time Ornstein-Uhlenbeck (OU) process~$\lbrace X(t)\colon t\geq0\rbrace$, expressed in its integral form as
In order to obtain tractable exit times, we consider a continuous time Ornstein-Uhlenbeck (OU) process~$\lbrace X(t)\colon t\geq0\rbrace$, whose solution at time~$t$ is expressed as
\begin{equation}
X(t) = e^{\mathcal{A}t}X(0) + \int_{0}^{t} e^{\mathcal{A}(t-s)}\mathcal{Q}\mathrm{d}W(s),
\label{eq:XOUSoln}
\end{equation}
with $\mathcal{A}$ the closed-loop Hurwitz matrix obtained from~(\ref{eq:ProcessControl}), $\mathcal{Q}$ the positive definite state covariance matrix, and $W(s)\in\mathbb{R}^{n_{x}}$~a Wiener process. For ease of presentation, we omit the agent subscript~$i$ for the derivation.

Similarly, the deterministic predictors~(\ref{eq:PredictorControl}) are given by
\begin{equation}
\hat{X}(t) = e^{\mathcal{A}t}X(0).
\label{eq:XhatOUSoln}
\end{equation}
%The error process $Z(t) \coloneqq X(t) - \hat{X}(t)$, which is itself an OU process starting at zero and is reset to zero when the state is communicated, will exit any bounded domain almost surely.
The error process $Z(t) \coloneqq X(t) - \hat{X}(t)$, which is itself an OU process starting at zero and is reset to zero when the state is communicated, will exit any bounded domain with non-zero probability due to the unboundedness of Gaussian distributions.
If we consider a threshold domain $\mathcal{D} = \lbrace Z(t) : \left\|Z(t)\right\|_{2} < \delta\rbrace$, the exit time~$\tau$ representing the first time the error process exits domain~$\mathcal{D}$, is defined as
%\db{will exit any bounded domain with non-zero probability would be an alternative formulation, due to the unboundedness of the Gaussian distribution if you want to make sure that really everyone gets the point}
\begin{equation}
\tau \coloneqq \mathrm{inf}\lbrace t \in \mathbb{R} : \left\|Z(t)\right\|_{2} \geq \delta\rbrace.
\label{eq:FET}
\end{equation}
We are now able to define the exit probability
\begin{equation}
H(z,t^{*}) \coloneqq \mathrm{P}(\tau < t^{*} \mid Z(0) = z),
\label{eq:ExitProbability}
\end{equation}
as the probability that the error process, starting at $z$, exits domain~$\mathcal{D}$ before the time~$t^{*}$. For the OU process $Z(t)$, the following lemma results in the exit probability.
%\begin{lem}
%Consider the initial boundary value problem
%\begin{align}
%\frac{\partial S}{\partial t^{*}} = \frac{\partial S}{\partial z}\mathcal{A}z + \frac{1}{2}\mathrm{tr}\left[ \mathcal{Q}^{T}\frac{\partial^{2} S}{\partial z^{2}}\mathcal{Q}\right]  \qquad &\mathrm{for} \ z \in \mathcal{D},\nonumber \\
%S(z,t) = 0 \qquad &\mathrm{for} \ z \in \partial \mathcal{D},\label{eq:aut_PDEeqn_OU}\\
%S(z,0) = 1 \qquad &\mathrm{for} \ z \in \mathcal{D}.\nonumber
%\end{align}
%with $t^{*}$$\,>\,$$0$, $\mathrm{tr}(\cdot)$ the trace of a matrix, and whose solution $S$$\,\coloneqq\,$$S(z,t^{*})$ represents the probability that an OU process starting at $z$ remains inside region $\mathcal{D}$ until time $t^{*}$. Then~$H(z,t^{*}) = \mathrm{P}(\tau < t^{*}\,|\,Z(0) = z) = 1 - S(x,t^{*})$.
%\end{lem}
\begin{lem}
	Consider the initial boundary value problem
	\begin{align}
	\frac{\partial S}{\partial t^{*}} = \frac{\partial S}{\partial z}\mathcal{A}z + \frac{1}{2}\mathrm{tr}\left[ \mathcal{Q}^{T}\frac{\partial^{2} S}{\partial z^{2}}\mathcal{Q}\right]  \qquad &\mathrm{for} \ z \in \mathcal{D},\nonumber \\
	S(z,t^{*}) = 0 \qquad &\mathrm{for} \ z \in \partial \mathcal{D},\label{eq:aut_PDEeqn_OU}\\
	S(z,0) = 1 \qquad &\mathrm{for} \ z \in \mathcal{D}.\nonumber
	\end{align}
	with $t^{*}\!>\!0$ and $\mathrm{tr}(\cdot)$ the trace of a matrix. Then the solution $S\!\coloneqq\!S(z,t^{*})$ denotes the survival probability of an OU process, starting at~$z$ and remaining inside region~$\mathcal{D}$ until time~$t^{*}$, and the resulting exit probability $H(z,t^{*})=\mathrm{P}(\tau<t^{*}\,|\,Z(0)=z)=1-S(z,t^{*})$.
\end{lem}
\begin{pf}
It\^{o}'s formula applied to an OU process results in the partial differential equation (\ref{eq:aut_PDEeqn_OU}), whose solution represents the survival probability $\mathrm{P}(\tau \geq t^{*}\,|\,Z(0) = z)$, as given in \cite{StochSystems}. The exit probability~$H(z,t^{*})$ then results from the probability complement law.\hfill$\blacksquare$
%	\db{The way you have it now the lemma basically only states that you can derive the exit probability as 1 - survival probability, what is not really surprising. If you formulate the lemma a bit differently, you can state as a first result that S(z,t) is the solution of this PDE and as a second result that you can use its complement to get the exit probability. In the proof you in the end anyway show both. Then I think it is fine to keep it as a lemma with proof.}
\end{pf}
%Analytic solutions of~(\ref{eq:aut_PDEeqn_OU}) have yet to be discovered for high dimensional processes. 
%Numerical approximations are therefore required, many of which pose computational and implementation challenges as the dimension $n_{x}$ increases. 
%Monte Carlo simulations present a simple alternative to these complex implementations, without suffering too much in accuracy, and are used in this paper to obtain the function $H(z,t^{*})$. No matter the computation method, solving (\ref{eq:aut_PDEeqn_OU}) at each time step proves to be too computationally demanding, particularly if the framework should be implemented on real-time embedded hardware. Therefore, solutions are computed offline and stored in lookup tables for quick runtime access.

%Ideally, solving (9) to evaluate $H(z,t^{*})$ at each discrete time step
To compute $H(z,t^{*})$, Monte Carlo simulations of the process $Z(t)$ are performed and the empirical probability of exiting domain $\mathcal{D}$ is obtained.
Approximating (\ref{eq:aut_PDEeqn_OU}) at each time step proves to be too computationally demanding, particularly if the framework is to be implemented on embedded hardware. Thus, solutions are computed offline and stored in lookup tables for quick runtime access.
 
%\db{This paragraph is nice the way you have it. If we get into troubles with space in the end, this might be a place to shorten. We can also just say that we use MC instead of discussing other alternatives. But for the moment I would leave it as it is, just keep it in mind for the final iterations}
\subsection{$M$-step communication probability}
\label{sec:PredHorizon}
%\begin{figure}[h!]
%	\begin{center}
%		\includegraphics[width=8.4cm]{img/pdf/prediction_horizon.pdf}
%		\caption{} 
%		\label{fig:prediction horizon}
%	\end{center}
%\end{figure}

%Now that we have a way of computing the exit probability $H(z,t^{*})$,
Utilizing the exit probability $H(z,t^{*})$, we can now derive the probability of communicating at an $M$-step time horizon, taking into account communication occurrences within said horizon. This probability is defined as the $M$-step communication probability $\mathrm{P}(C_{t+M\Delta t})$, with $\Delta t$ the time step size.
%Now, we derive the $M$-step communication probability $\mathrm{P}(C_{t+M\Delta t})$, utilizing the computed exit probability and taking into account communication occurrences within the prediction horizon~$M$.
Let $C_{t}\!\in\!\lbrace0,1\rbrace \ \forall \ t\!>\!0$ be probabilistic events denoting communication at time $t$, and whose outcomes represent negative and positive communication decisions, respectively. The $M$-step communication probability is then expressed as
\vspace{1mm}
\begin{align}
%\mathrm{P}(C_{t+M\Delta t}) 
%& = \sum_{\substack{C_{t+(M-1)\Delta t},\\\dots,C_{t}}} \mathrm{P}(C_{t+M\Delta t},\dots,C_{t}) \label{eq:P(C))chain} \\
%&= \sum_{\substack{C_{t+(M-1)\Delta t},\\\dots,C_{t}}} \begin{bmatrix}
%\prod\limits_{m=0}^{M} \mathrm{P}\begin{pmatrix}C_{t+m\Delta t} \Biggm\vert \bigcap\limits_{l=0}^{m-1} C_{t+l\Delta t} \end{pmatrix}\end{bmatrix}. \nonumber
%\mathrm{P}(C_{t+M\Delta t}) 
%& = \sum_{\substack{C_{t+(M-1)\Delta t},\\\dots,C_{t}}} \mathrm{P}(C_{t+M\Delta t},\dots,C_{t}) \label{eq:P(C))chain} \\
%&= \sum_{\substack{C_{t+(M-1)\Delta t},\\\dots,C_{t}}} \prod\limits_{m=0}^{M} \mathrm{P}\begin{pmatrix}C_{t+m\Delta t} \Biggm\vert \bigcap\limits_{l=0}^{m-1} C_{t+l\Delta t} \end{pmatrix}.
\mathrm{P}(C_{t+M\Delta t}) 
& = \sum_{C_{t+(M\!-\!1)\Delta t},\dots,C_{t}} \mathrm{P}(C_{t+M\Delta t},\dots,C_{t}) \label{eq:P(C))chain} \\
&= \sum_{\substack{C_{t+(M\!-\!1)\Delta t},\\\dots,C_{t}}} \prod\limits_{m=0}^{M} \mathrm{P}\begin{pmatrix}C_{t+m\Delta t} \Biggm\vert \bigcap\limits_{l=0}^{m-1} C_{t+l\Delta t} \end{pmatrix}. \nonumber
\end{align}

%The Markov property is exploited to render the problem tractable, by expressing the joint probability $\mathrm{P}(C_{t+M\Delta t},\dots,C_{t})$ in terms of current and future communication events, and not past events.
%which exploits the Markov property to express the joint probability $\mathrm{P}(C_{t+M\Delta t},\dots,C_{t})$ with dependency on current and future communication events, but not on past events.
To render the problem tractable, the joint probability $\mathrm{P}(C_{t+M\Delta t},\dots,C_{t})$ in (\ref{eq:P(C))chain}) is assumed to depend on current and future communication events, but not on past events.
%By exploiting the Markov property, the joint probability $\mathrm{P}(C_{t+M\Delta t},\dots,C_{t})$ becomes dependent on current and future communication events, but not on past events. 
For instance, the communication probability for a prediction horizon $M=2$ may be expanded as
\vspace{1mm}
\begin{align}
\mathrm{P}(C_{t+2\Delta t}) &= \mathrm{P}(C_{t+2\Delta t} \mid C_{t+\Delta t},C_{t})\mathrm{P}(C_{t+\Delta t} \mid C_{t})\mathrm{P}(C_{t}) \nonumber \\&+ \, \mathrm{P}(C_{t+2\Delta t} \mid C_{t+\Delta t},\bar{C}_{t})\mathrm{P}(C_{t+\Delta t} \mid \bar{C}_{t})\mathrm{P}(\bar{C}_{t}) \nonumber \\
&+ \, \mathrm{P}(C_{t+2\Delta t} \mid \bar{C}_{t+\Delta t},C_{t})\mathrm{P}(\bar{C}_{t+\Delta t} \mid C_{t})\mathrm{P}(C_{t}) \nonumber \\ &+ \, \mathrm{P}(C_{t+2\Delta t} \mid \bar{C}_{t+\Delta t},\bar{C}_{t})\mathrm{P}(\bar{C}_{t+\Delta t} \mid \bar{C}_{t})\mathrm{P}(\bar{C}_{t}), \nonumber
\end{align}
where $\lbrace\bar{C}_{t},{C}_{t}\rbrace$ are used above as shorthand to denote the outcomes $\lbrace{C}_{t}\!=\!0, {C}_{t}\!=\!1\rbrace$, respectively.
%where $(\bar{ \ \cdot \ })$ denotes the set complement.
Since the PT framework may only allocate slots to a subset of agents, the actual probability that an agent will communicate at a future time step within the horizon is unknown at the current time. The following assumption is then required to estimate the conditional probabilities in (\ref{eq:P(C))chain}).
\begin{assum}
	At the current time $t$, the probability of communicating at times $t+m\Delta t, \ \forall \ m = \lbrace 1,\dots,M\rbrace$, given preceding communication events, is approximated by 
	\begin{equation}
	\mathrm{P}(C_{t+m\Delta t} \mid C_{t+(m\!-\!1)\Delta t},\dots,C_{t}) \approx H(z,t^{*}),
	\end{equation}
	where the error $z$ and time to exit $t^{*}$ are obtained based on the outcomes of the conditional communication events.
	\label{ass:PCommEstimate}
\end{assum}
\vspace{-1em}
A few examples of how to obtain $z$ and $t^{*}$ are now presented. If we consider a horizon $M=3$, $\mathrm{P}(C_{t+3\Delta t})$ is expanded into several conditional probabilities, one of which being $\mathrm{P}(C_{t+3\Delta t} \mid C_{t+2\Delta t},\bar{C}_{t+\Delta t},C_{t})$. 
Since communication resets the error to $z=0$, and communications occur at times $t$ and $t+2\Delta t$, it is only necessary to consider the latest communication event at $t+2\Delta t$ when applying Assumption~\ref{ass:PCommEstimate}.
%Since communication resets the error to $z=0$, and communications occur at times $t$ and $t+2\Delta t$, it is only necessary to consider the latest communication event at $t+2\Delta t$ when estimating the conditional probability.
A single time step remains between the last positive communication event and $t+3\Delta t$, resulting in $\mathrm{P}(C_{t+3\Delta t} \mid C_{t+2\Delta t},\bar{C}_{t+\Delta t},C_{t}) \approx H(0,\Delta t)$. 
Similarly, if we consider $\mathrm{P}(C_{t+2\Delta t} \mid \bar{C}_{t+\Delta t},\bar{C}_{t})$ where no communications occur within the horizon, the conditional probability is approximated by $H(z,2\Delta t)$, where $z$ is the current error.
%Otherwise, if no positive communication events occur within the horizon, the error 

All agents compute $\mathrm{P}(C_{t+M\Delta t})$ and communicate it periodically to the scheduler, which utilizes these probabilities to prioritize allocation.
Prior to communicating the probability, each agent casts said probability into a 1-byte integer, thereby reducing bandwidth consumption, \eg if $\mathrm{P}(C_{t+M\Delta t})\!=\!0.48173$, send $\mathrm{P}(C_{t+M\Delta t})\!=\!48$.
%Prior to communicating the probability, each agent applies a casting operation to convert said probability into a 1-byte integer, thereby reducing bandwidth consumption, \eg if $\mathrm{P}(C_{t+M\Delta t}) = 0.48173$, send $\mathrm{P}(C_{t+M\Delta t}) = 48$.
%\jm{value, because 1-byte is technically not an integer anymore? this is a nitpick but might be important to change}
%The $M$-step communication probability sent to the network manager, by all agents, serves as the priority measure for resource allocation. We cast the computed probabilities into integers, thereby reducing bandwidth consumption.\jm{fix}
An extension to further reduce communication may also be applied:
\vspace{1mm}
\begin{rem}
The PT framework can be extended by applying a lower bound $p$ on the probabilities sent to the scheduler by agent~$i$. Formally, this is expressed as 
\begin{equation}
\mathrm{at\ time\ k:} \ \ \mathrm{send \ P}_{k}^{i}(C_{k+M}) \iff \mathrm{P}_{k}^{i}(C_{k+M}) > p.
\end{equation}
This extension allows the scheduler to disregard agents that are unlikely to require communication in $M$ steps, thereby further reducing the network utilization.
\label{rem:LowerBoundExtension}
\end{rem}
%\begin{rem}
%\textit{Another possible extension of the framework is to compute and send the communication probabilities periodically with period $R>1$. The sparsity of resource allocation and communication decisions resulting from sending the probability at lower rates reduces the network utilization at the cost of an increase in control error.}\jm{I'm leaning towards removing it.}
%\label{rem:PPTExtension}
%\end{rem}

\subsection{Communication decisions}
\label{sec:NetMan}
The network scheduler proposed herein, allocates resources at time $k+M$ by ranking agents according to their probabilities, \ie the~$K$ agents with the largest probabilities receive a slot. This is represented as
%For agents that send their probability at time $k$, the \textit{Scheduler} in \figref{fig:distributed_PT_control} allocates available resources $M$ steps in advance.
%The resource allocation decisions $\gamma_{s,k+M}^{i}$\jm{again, i think this should be k instead of k+M?}\db{See my previous comment. It must become clear that the decision for time k depends on information sent at time k-M, or the decision at k+M depends on information sent at k. Which way round doesn't matter too much I'd say, just should be consistent throughout the paper}, are given by
\begin{align}
\mathrm{at\ time\ k:}\quad \gamma_{s,k+M}^{i} = 1 \iff \mathrm{P}_{k}^{i}(C_{k+M}) \in \mathcal{P}_{k},
\end{align}
where $\mathcal{P}_{k}$ is the set of $K$ largest probabilities sent ot the scheduler at time $k$.
The \textit{Trigger}, located in each agent, then takes communication decisions $\gamma_{c,k}^{i}$, based on the slot allocation and the estimation error. This is given by
\begin{align}
\mathrm{at\ time\ k:}\quad \gamma_{c,k}^{i} = 1 \iff \gamma_{s,k}^{i} = 1 \land \left\|z_{k}^{i}\right\|_{2} \geq c\delta,
\label{eq:Trigger}
\end{align}
where $c\!\in\!\lbrack0,1\rbrack$ is a design parameter and $\land$ denotes the logical AND operator.
If $c\!=\!0$, communication occurs whenever the agent is allocated a slot.
%Increasing $c$ results in utilization improvements but presents some risk since the next time a slot will be allocated is unknown.
Increasing $c$ improves the utilization, but has a negative impact on performance.
%As $c$ increases, the resulting utilization improvements present some risk, since the next time a slot will be allocated is unknown.

\section{Autonomous Vehicle Platooning}
\label{sec:CACC}
%In this section, we present an application of the PT framework on CACC-equipped autonomous vehicle platoons\db{we apply the PT framework to vehicle platooning would probably save you a line}.
In this section, we apply the PT framework to CACC-equipped autonomous vehicle platoons. Through intervehicle communication and radar-sensed information, CACC improves traffic throughput and reduces fuel consumption by maintaining small intervehicle distances, as shown in \cite{besselink2016cyber,ET_platoon}.
%Here, the time gap refers to the time required for a vehicle to cover the distance between its position and the position of the preceding vehicle.\jm{is this necessary?}

\subsection{Control design}
\label{sec:CACC_ControlDesign}

The following control design, vehicle model, and control law are taken from \cite{ET_platoon}. CACC systems obey two control objectives: 1) each vehicle~$i$, of length~$L_{i}$, must follow the preceding vehicle~$i\!-\!1$ with a desired distance $d_{r,i}(t)$, known as \emph{individual vehicle stability}, and, 2) disturbances are attenuated along the platoon, known as~\emph{string stability}. Consider the vehicle model:
\begin{subequations}
	\begin{align}
%	\dot{s}_{i}(t) &= v_{i}(t), \label{eq:VehiclePos} \\
	\dot{v}_{i}(t) &= a_{i}(t), \label{eq:VehicleVel} \\ 
	\dot{a}_{i}(t) &= -\frac{1}{\tau_{i}}a_{i}(t) + \frac{1}{\tau_{i}}\alpha_{i}(t), \label{eq:VehicleAcc} \\
	\dot{e}_{i}(t) &= v_{i-1}(t) - v_{i}(t) - h_{i}a_{i}(t), \label{eq:VehicleSpaceError} \\
	\dot{\alpha}_{i}(t) &= -\frac{1}{h_{i}}\alpha_{i}(t) + \frac{1}{h_{i}}u_{i}(t), \label{eq:VehicleDesAcc}
	\end{align}
\end{subequations}
where $v_{i}(t)$, $a_{i}(t)$, $\alpha_{i}(t)$, and $u_{i}(t)$ are the velocity, acceleration, desired acceleration, and control input of vehicle $i$ at time $t$, respectively. The spacing error between vehicle~$i$ and $i\!-\!1$ is denoted by $e_{i}(t)$, $\tau_{i}\!>\!0$ is a time constant representing the engine dynamics, and $h_{i}\!\coloneqq\!(d_{r,i}(t)\!-\!r_{i})/v_{i}(t)$ is the time gap, with $r_{i}$ the standstill distance of vehicle $i$.

The CACC system considered herein follows a predecessor-following communication topology (\ie each vehicle $i$ receives communication from its preceding vehicle $i\!-\!1$), with the state of vehicle $i$ defined as $x_{i} \coloneqq \lbrack e_{i}, \dot{e}_{i}, \ddot{e}_{i}, \alpha_{i}\rbrack^{\mathrm{T}}$. The cooperative control law is given by
%If we define the state of vehicle $i$ as $x_{i} \coloneqq \begin{bmatrix} e_{i} & \dot{e}_{i} & \ddot{e}_{i} & \alpha_{i}\end{bmatrix}^{\mathrm{T}}$, a cooperative control law is given by
\begin{equation}
u_{i}(t) = F\begin{bmatrix} e_{i} & \dot{e}_{i} & \ddot{e}_{i} \end{bmatrix}^{\mathrm{T}} + \tilde{\alpha}_{i-1}(t),
\label{eq:CACC_ControlLaw}
\end{equation}
with $F = \begin{bmatrix} k_{p} & k_{d} & k_{dd}\end{bmatrix}$ the gain matrix and $\tilde{\alpha}_{i-1}(t)$ the most recent information received by vehicle~$i$ regarding $\alpha_{i-1}$ .
%with $F = \begin{bmatrix} k_{p} & k_{d} & k_{dd}\end{bmatrix}$ the gain matrix and $\tilde{\alpha}_{i\!-\!1}(t)$ the desired acceleration of vehicle $i-1$ perceived\jm{fix?} by vehicle~$i$.
%Due to the communication topology, each vehicle only needs to run predictions of the preceding vehicle $i-1$. 
%The control law~(\ref{eq:ControlInput}) may then be expressed as 
%The CACC system considered herein follows a predecessor-following communication topology, therefore, vehicle~$i$ only needs to run predictions of the preceding vehicle $i-1$.
%In the PT framework, vehicle~$i$ runs predictions of the preceding vehicle $i-1$, and the last state variable of $\hat{x}_{i-1}(t)$ provides an estimate of $\tilde{\alpha}_{i-1}(t)$. 
%The PT control law~(\ref{eq:ControlInput}) is then given by
%\begin{equation}
%u_{k}^{i} = F\hat{x}_{1:3,k}^{i} + \hat{x}_{4,k}^{i-1},
%\label{eq:PT_CACC_ControlLaw}
%\end{equation}
%with $\hat{x}_{1:3,k}^{i}$ the first three, and $\hat{x}_{4,k}^{i-1}$  the last state variable(s) of the predicted state of vehicles $i$ and $i-1$, respectively.

\subsection{PT and ET designs}
\label{sec:ETDesign}

Due to the communication topology, each vehicle only needs to run predictions of the preceding vehicle $i\!-\!1$, and control law~(\ref{eq:ControlInput}) may be expressed as
%The control law~(\ref{eq:ControlInput}) may then be expressed as
\begin{equation}
u_{k}^{i} = F\hat{x}_{1:3,k}^{i} + \hat{x}_{4,k}^{i-1},
\label{eq:PT_CACC_ControlLaw}
\end{equation}
$\mathrm{with}\,\,\hat{x}_{1:3,k}^{i}\mathrm{\,\,the\,first\,three\,and\,}\,\,\hat{x}_{4,k}^{i-1}\mathrm{\,\,the\,last\,state\,variable(s)}$ of the predicted state of vehicles $i$ and $i\!-\!1$,~respectively.

The PT is implemented with parameters $M\!=\!2$,~$c\!=\!0.75$, and extended with $p\!=\!0.2$ as defined in Remark~\ref{rem:LowerBoundExtension}, while~the exit probability is computed from \SI{10000}{} Monte Carlo simulations.
%The probabilistic measure used to allocate resources is then obtained by using the exit probability computed from $10 000$ Monte Carlo simulations. 
The ET designs are implemented for comparison, by setting the PT horizon to $M\!=\!0$. Since ETs decide instantaneously whether communication is necessary, the scheduler must preallocate resources independently from the triggering decisions. 
We compare the approach developed herein to the following two methods: ET1) allocation to $K$ random agents at each time step $k$, as in \cite{Molin}, and, ET2) allocation according to an error-dependent priority measure, as in \cite{ErrordependentDS}.
%2) allocation to $K$ agents~$\forall \ k$, denoted ET2.
%Due to the resource constraint considered in this paper, allocation of said resources is critical to the string stability of platoons. 

\subsection{Results}
\label{sec:CACCResults}
%\db{You have a lot of parameters here that need to be introduced. One possibility would be to do it in a table has you have it in your thesis, but this probably needs more space. 
%Another possibility would be to do it the same way we have it in Sec. VI in the IoT paper.
%Maybe just take a look, that's a very compact form that should save you some space and I think it's easier to read than with all the connecting words in between}
%\begin{table}[htb]
%	\begin{center}
%		\caption{Simulation parameters.}
%		\label{tab:CACCSim_Params}
%		\begin{tabular}{||l || c || l | c ||}
%			\hline
%			CACC & Value & PT & Value \\ [0.5ex] 
%			\hline
%			$L_{i}$ & \SI{4}{\meter} & $T$ & \SI{120}{\second} \\
%			\hline
%			$r_{i}$ & \SI{2.5}{\meter} & $\Delta t$ & \SI{0.01}{\second} \\
%			\hline
%			$\tau_{i}$ & \SI{0.1}{\second} & $K$ & 20 \\
%			\hline
%			$h_{i}$ & \SI{0.7}{\second} & $M$ & $2$ \\
%			\hline
%			$k_{p_{i}}$ & $0.2$ & $\delta$ & $0.01$ \\	
%			\hline
%			$k_{d_{i}}$ & $0.7$ & $p$ & $0.2$  \\	
%			\hline
%			$k_{dd_{i}}$ & $0$ & $c$ & $0.75$ \\			
%			\hline
%		\end{tabular}
%	\end{center}
%\end{table}
%\begin{table}[htb]
%	\begin{center}
%		\caption{Simulation parameters.}
%		\label{tab:CACCSim_Params}
%		\begin{tabular}{||l || c || l | c ||}
%			\hline
%			CACC & Value & PT & Value \\ [0.5ex] 
%			\hline
%			$L_{i}$ & \SI{4}{\meter} & $M$ & $2$ \\
%			\hline
%			$r_{i}$ & \SI{2.5}{\meter} & $\delta$ & $0.01$ \\
%			\hline
%			$\tau_{i}$ & \SI{0.1}{\second} & $p$ & $0.2$ \\
%			\hline
%			$h_{i}$ & \SI{0.7}{\second} & $c$ & $0.75$ \\			
%			\hline
%		\end{tabular}
%	\end{center}
%\end{table}
Simulations of a CACC-equipped vehicle fleet of varying system size $N$, divided evenly over an $l=5$ lane~highway, are performed for $T=\SI{120}{\second}$. Each lane consists~of~a single platoon and follows a virtual reference vehicle with velocity $v_{\mathrm{ref}}$. The simulations are analyzed with:
\vspace{-0.3em}
\begin{example}
	$K\!=\!20$ slots, $\Delta t\!=\!\SI{0.01}{\second}$, $L_{i}\!=\!\SI{4}{\meter}$, $r_{i}\!=\!\SI{2.5}{\meter}$, $\tau_{i}\!=\!\SI{0.01}{\second}$, $h_{i}\!=\!\SI{0.7}{\second}$, $\Sigma_{w}\!=\!9\!\cdot\!10^{-6}I_{4}$, and $\delta_{i}\!=\!0.01 \ \forall \ i\!\in\!\Omega$.
	\label{ex:CACC}
\end{example}
%The simulations run for $T=\SI{120}{\second}$, with a step size $\Delta t=\SI{0.01}{\second}$ and the system communicates over a network with $K=20$ communication slots. The CACC and PT parameters used in the simulation are presented in \tabref{tab:CACCSim_Params}, 

All vehicles adopt controller~(\ref{eq:PT_CACC_ControlLaw}) with gains set to $k_{p}\!=\!0.2$, $k_{d}\!=\!0.7$, and $k_{dd}\!=\!0$, satisfying the stability criteria needed to provide individual vehicle stability as in \cite{ET_platoon}.
%All vehicles adopt controller~(\ref{eq:PT_CACC_ControlLaw}) with gains set to $k_{p}=0.2$, $k_{d}=0.7$, and $k_{dd}=0$. In~\cite{CACC,ET_platoon}, the above gains are shown to satisfy the stability conditions and provide individual vehicle stability.
%The CACC parameters are set to $L_{i}=\SI{4}{\meter}$, $r_{i}=\SI{2.5}{\meter}$, $\tau_{i}=\SI{0.01}{\second}$, $h_{i}=\SI{0.7}{\second}$, and
%The PT parameters are set to: $M=2$, $\delta=0.01$, $c = 0.75$ and the framework is extended with the lower bound $p=0.2$, as presented in Remark \ref{rem:LowerBoundExtension}.
The performance of the CACC system is evaluated using the control error of vehicle~$i$ at time~$k$, defined as
\begin{equation}
\varepsilon_{k}^{i} \coloneqq \begin{bmatrix}v_{k}^{i} - v_{\mathrm{ref}} \\ d_{k}^{i} - d_{r,k}^{i} \end{bmatrix},
\label{eq:CACCControlError}
\end{equation}
with $d_{k}^{i}$ the radar-sensed intervehicle distance and $d_{r,k}^{i}$ the desired intervehicle distance computed as in~\cite{ET_platoon}, dependent on the time gap~$h_{i}$.
%The performance metrics can now be defined.
For the comparison, we consider the mean control error norm $\bar{E}$, and mean network utilization $\bar{U}$ as performance metrics, defined as
\begin{equation}
\bar{E} \coloneqq \frac{1}{N\mathcal{T}} \sum\limits_{i=1}^{N}\sum\limits_{k=1}^{\mathcal{T}} \left\|\varepsilon_{k}^{i}\right\|_{2}, \qquad
\bar{U} \coloneqq \frac{1}{\mathcal{T}} \sum\limits_{k=1}^{\mathcal{T}} U_{k}. \nonumber
\end{equation}
The number of time steps $\mathcal{T} = T/\Delta t$, while $\varepsilon_{k}^{i}$ is the control error of agent $i$, and $U_{k}$ is the network utilization  at time $k$, defined in~(\ref{eq:CACCControlError}) and~(\ref{eq:Utilization}), respectively.
%\begin{defn}
%	Consider a multi-agent simulation with runtime $T$ and time step size $\Delta t$. The mean control error norm $\bar{E}$, and mean network utilization $\bar{U}$, are given by
%	\begin{equation}
%	\bar{E} \coloneqq \frac{1}{N\mathcal{T}} \sum\limits_{i=1}^{N}\sum\limits_{k=1}^{\mathcal{T}} \left\|\varepsilon_{k}^{i}\right\|_{2}, \qquad
%	\bar{U} \coloneqq \frac{1}{\mathcal{T}} \sum\limits_{k=1}^{\mathcal{T}} U_{k}, \nonumber
%	\end{equation}
%	with $\mathcal{T} = T/\Delta t$ the number of time steps, $\varepsilon_{k}^{i}$ the control error of agent $i$, and $U_{k}$ the network utilization  at time $k$, defined in~(\ref{eq:CACCControlError}) and~(\ref{eq:Utilization}), respectively.
%	\label{def:PerfMetrics}
%\end{defn}

\begin{figure}[htb]
	\begin{center}
		\includegraphics[width=8.2cm]{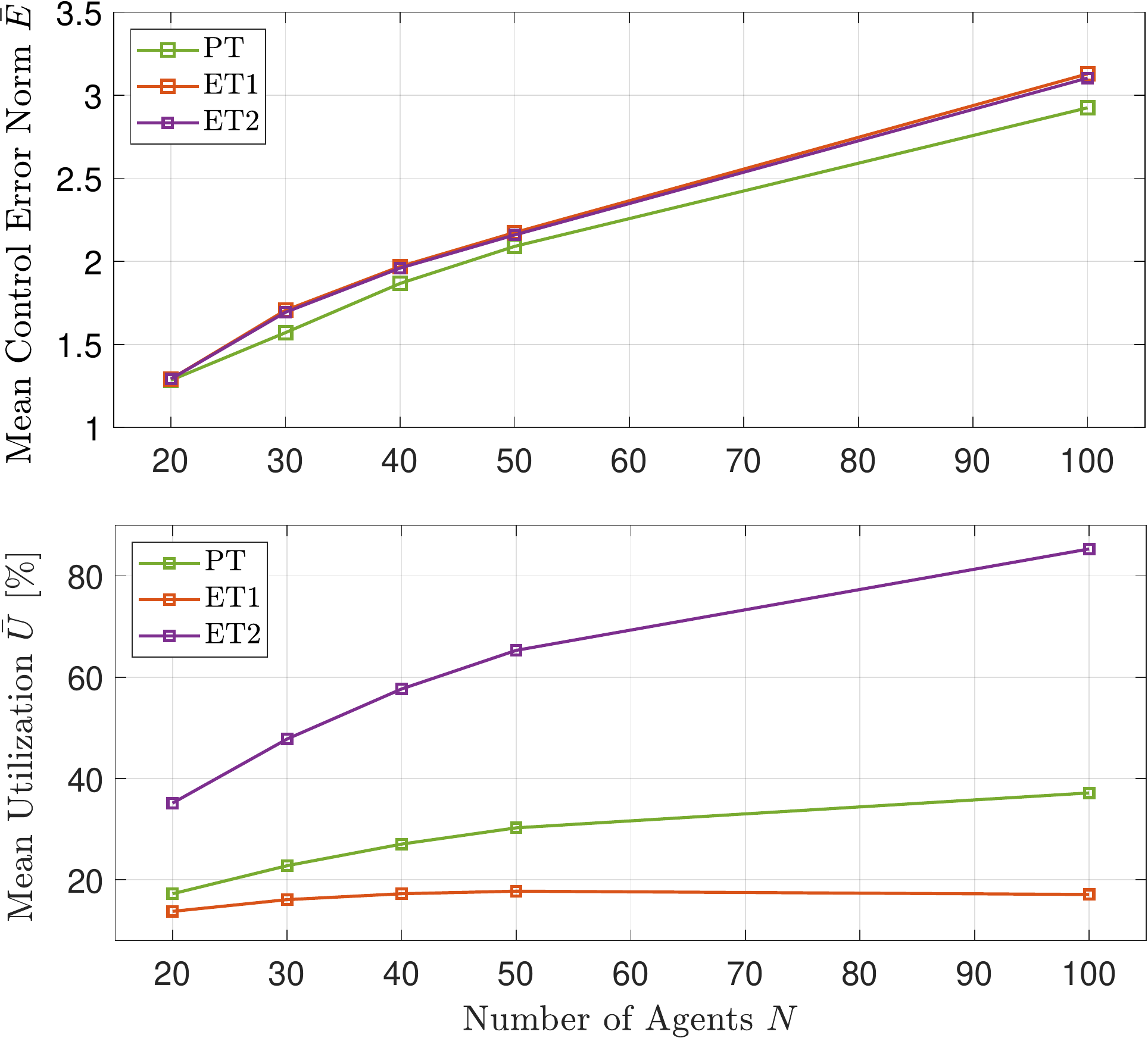}    % The printed column width is 8.4 cm.
		\caption{CACC simulations (\cf~\secref{sec:CACC}) for varying system size $N$, comparing the PT and ET designs. TOP: Mean control error norm $\bar{E}$. BOTTOM: Mean utilization $\bar{U}$.}
%		\caption{(a) Mean control error norm $\bar{E}$, and (b) mean network utilization $\bar{U}$ for varying system size $N$, using the ET and PT CACC simulations.}
		\label{fig:CACC_TrigComp}
	\end{center}
\end{figure}

For varying system size $N$, the tradeoff between mean control error norm $\bar{E}$ and mean network utilization $\bar{U}$ is depicted in~\figref{fig:CACC_TrigComp}.
Since the ET1 design does not require sending information to a scheduler, as slots are allocated randomly (\ie $b\!=\!0$ in~(\ref{eq:Utilization})), we take its utilization curve as a lower bound, achieving the lowest utilization while expectedly resulting in the worst performance.
%since resource allocation occurs without the need for extra communication.
%We observe that while the various allocation methods result in similar performance, due to the point-to-point communication topology (\ie a vehicle sends and receives information to/from one other vehicle), 
We observe that while the ET designs result in similar performance, the PT is able to reduce $\bar{E}$ by allocating resources in advance.
The PT is also able to significantly reduce $\bar{U}$, compared to the ET2 design.
This is because with the PT, $N_{p,k}\!\leq\!N$ agents send a $b\!=\!1$ byte (integer) probability to the scheduler, whereas with the ET2 design $N_{p,k}\!=\!N$ agents send a $b\!=\!4$ byte (float) error norm.
%We also see that with the PT, more resources remain available (\ie a network utilization below~100\%) to either be shut off to save more energy, or allocated for other purposes.
The utilization improvements resulting from the PT also allow us to either increase energy savings by shutting off unused resources, or allocating them for other purposes.
%It can be seen that the PT is able to significantly reduce $\bar{U}$, compared to the ET2 design. With the PT, $N_{p,k} \leq N$ agents send a 1-byte (integer) probability to the scheduler, whereas with the ET2 design $N_{p,k} = N$ agents send a 4-byte (float) error norm.
%The extended PT* design is also shown to further reduce utilization, compared to the regular PT, while maintaining the control error. This occurs since agents only send their probability when it is above the imposed lower bound $p$, \ie when agents have a high enough communication demand.
%The PT framework, on the other hand, results in lower $\bar{E}$ than the ET1 method, both of which are associated with the left y-axis. 
%As the system size increases, we also observe that the PT and ET1 designs begin to diverge, showing the PT designs effectiveness at handling situations when $N \gg K$.
%This result comes at the expense of increased network utilization as a result of communicating probabilities.
%However, the PT framework is able to minimize the amount of information exchanged over the network, and allow remaining resources to either be shut off to save energy or reallocated for other purposes.\jm{remove?}
\vspace{1.5mm}

\section{Cart-Pole Hardware Experiments}
\label{sec:CP}

To validate the framework, we consider an experimental multi-agent cart-pole system composed of a single self-built physical agent and several simulated agents, all of which are implemented on and communicate through a laptop running Matlab/Simulink.
The cart position~$s_{i}$, pole angle $\theta_{i}$, and their velocities, form the state $x_{i}\!=\!\lbrack s_{i}, \theta_{i}, \dot{s_{i}}, \dot{\theta_{i}}\rbrack^{\mathrm{T}}$, where $s$ and $\theta$ of the physical agent are measured through encoder sensors and $\dot{s}$ and $\dot{\theta}$ are obtained via finite differences.
%We obtain the physical agent model through least-square system identification experiments, during which a \SI{60}{\second} chirp signal excites the agent.
Identifying the physical agent model through least-squares estimation results in:
%The mean of ten identification experiments results in the model\db{This may be shortened: Identifying the real system through least-squares estimation results in or sth like that}:
\begin{equation}
A = \begin{bsmallmatrix*}[r] 1.0006 & {-0.0034} & 0.0076 & 0.0009 \\ 0.0098 & 0.9785 & 0.0041 & 0.0057 \\ 0.0231 & {-0.1186} & 0.9268 & 0.0366 \\ {-0.0790} & 0.2596 & {-0.1350} & 1.0500 \end{bsmallmatrix*}, \ B=\begin{bsmallmatrix} 0.0003 \\ 0.0002 \\ 0.0076 \\ 0.0160 \end{bsmallmatrix}.
\label{eq:sysID}
\end{equation}
The simulated agents are corrupted by process noise~$w_{k}^{i}$, as in~(\ref{eq:ProcessControl}), while the physical agent is corrupted by input noise~$\epsilon_{k}^{i} \sim \mathcal{N}(0,\Sigma_{\epsilon})$.
The experiments are run with:
\vspace{-0.3em}
\begin{example}
	\hspace{-1em}
	$N\!=\!10$,~$T\!=\!\SI{30}{\second}$,~$\Delta t\!=\!\SI{0.01}{\second}$,~$\Sigma_{w}\!=\!2.5\!\cdot\!10^{-5}I_{4}$, $\Sigma_{\epsilon}\!=\!10^{-6}I_{4}$, PT parameters: $M\!=\!2$, $c\!=\!0.5$, $\delta_{i}\!=\!0.02 \ \forall \ i\!\in\!\Omega$, and \SI{10000}{} Monte Carlo simulations.
	\label{ex:CP}
\end{example}
%The heterogeneous system of $N=10$ agents is run for $T=\SI{30}{\second}$, with a step size $\Delta t=\SI{0.01}{\second}$, and for varying number of communication slots $K$. The PT horizon is set to $M=2$, the error bound to $\delta=0.02$, and the noise covariances are given by $\Sigma_{w}= 2.5\cdot10^{-5}I_{4}$ and $\Sigma_{\epsilon}= 10^{-6}I_{4}$.
%The considered system is heterogeneous for the following reasons: 1) an \SI{89}{\gram} mass is added to the physical agents pole, whose model is obtain by averaging ten \SI{60}{\second} system identification experiments, while the simulated agents use the model provided in \cite{QuanserManual}, and 2) the physical agent is corrupted by input noise, while the simulated agents are corrupted by process noise
\vspace{-1em}
\subsection{Synchronization}
%First, a synchronization experiment is presented, providing a comprehensive example of a distributed system, since synchronization requires communication between all agents in order for their states to evolve in unison.
First, we run a synchronization experiment with the goal to synchronize the cart positions $s_{i}$ of all agents.
This synchronization problem provides a comprehensive example of a distributed system since communication occurs between all agents. Such systems require bus-like networks, some of which have been successfully used for feedback control, both in wired (\cite{FieldBus}) and wireless (\cite{WirelessFeedback}) settings.
%This problem provides a comprehensive example of a distributed system since communication between all agents is required.
A homogeneous system is considered, with the simulated agent models also given by~(\ref{eq:sysID}), and we adopt the control design from \cite{WirelessFeedback}, which is based on a linear quadratic regulator (LQR).
The LQR parameters~$Q_{i}$,~$R_{i}$, and~$Q_{\mathrm{sync}}$, represent positive definite weight matrices penalizing deviations from the control objective, high control inputs, and deviations from the synchronization objective, respectively.
As we aim to synchronize the cart positions, we select $Q_{\mathrm{sync}}\!=\!\mathrm{diag}([30,0,0,0])$, and choose $Q_{i}\!=\!\mathrm{diag}([0.75,4,0,0])$ and $R_{i}\!=\!0.05$ for stabilization, where $\mathrm{diag}(\cdot)$ denotes a block-diagonal matrix.
%To make the synchronization more challenging, we disturb one of the simulated agents with a sine signal.
To make the synchronization more challenging, a sinusoidal disturbance is applied to one of the simulated agents.
This agents cart position is used as the reference when computing the control error of each agent, \ie the difference between each agents position and the position of the disturbed agent.
The PT is compared to the ET allocation designs presented in~\secref{sec:ETDesign}, and to the PT extended with $p\!=\!0.2$ (Remark~\ref{rem:LowerBoundExtension}), denoted by PT*.

\begin{figure}[htb]
	\begin{center}
		\includegraphics[width=8.3cm]{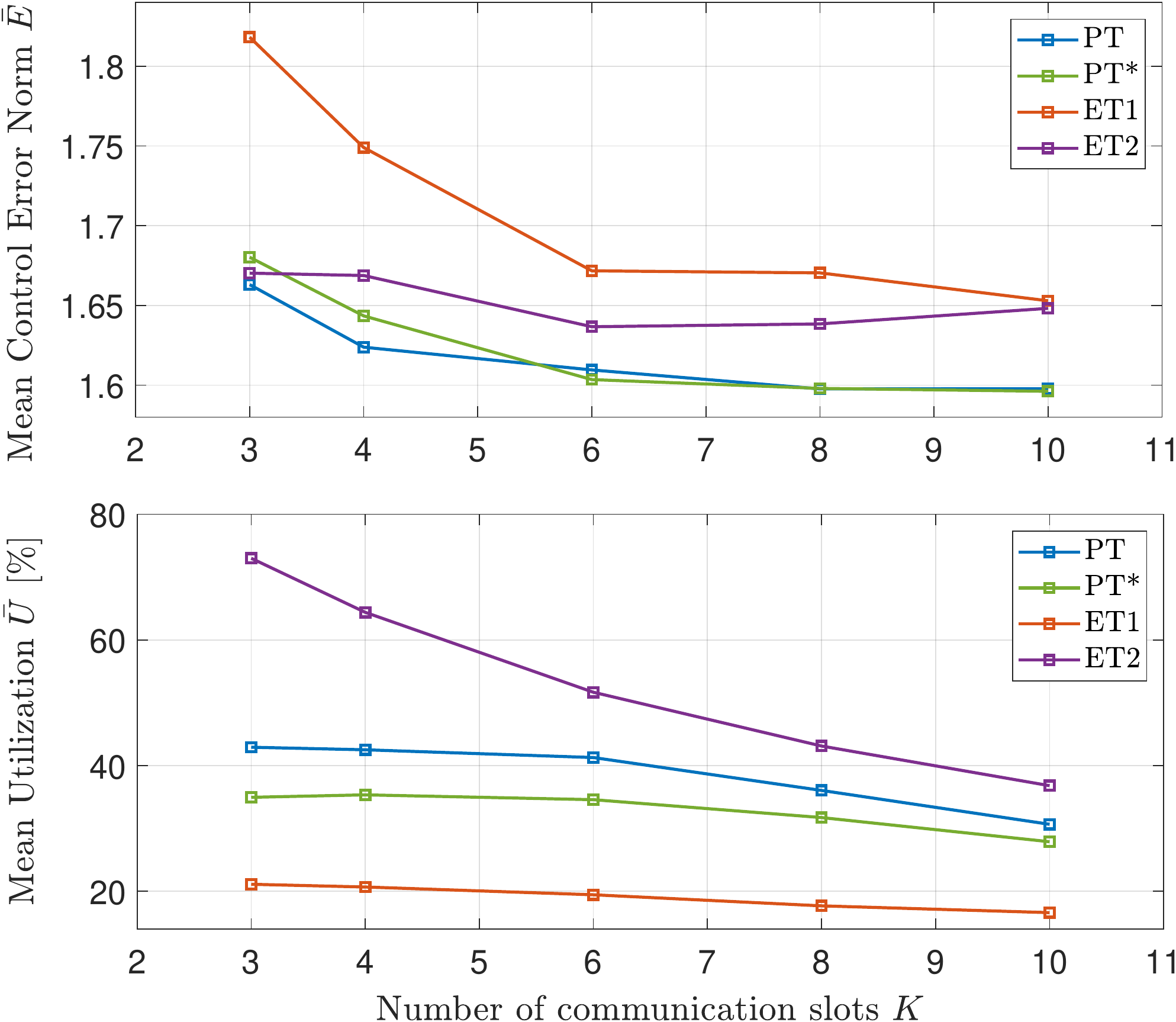}    % The printed column width is 8.4 cm.
		\caption{Cart-pole experiments  (\cf~\secref{sec:CP}) for varying number of communication slots $K$, comparing the various trigger designs. TOP: Mean control error norm $\bar{E}$. BOTTOM: Mean utilization $\bar{U}$.\newline}
%		\caption{(a) Mean control error norm $\bar{E}$, and (b) mean network utilization $\bar{U}$, for varying number of communication slots $K$, using the ET and PT on a multi-agent cart-pole system.}
		\label{fig:CP_sync_TrigComp}
	\end{center}
\end{figure}

For varying number of available communication slots~$K$, the tradeoff between mean control error norm~$\bar{E}$ and mean network utilization $\bar{U}$, is illustrated in~\figref{fig:CP_sync_TrigComp}.
For $K\!\leq\!2$, the systems become unstable due to insufficient resources, no matter the triggering method.
Similar to~\secref{sec:CACCResults}, the results show that the regular and extended PTs result in lower~$\bar{E}$ than both ET designs, and are once again able to reduce $\bar{U}$ compared to the ET2 design.
%With the PT, $N_{p,k} \leq N$ agents send a 1-byte (integer) probability to the scheduler, whereas with the ET2 design $N_{p,k} = N$ agents send a 4-byte (float) error norm.
The extended PT* design is also shown to further reduce utilization, compared to the regular PT, while maintaining the control error. This occurs since agents only send their probability when it is above the imposed lower bound $p$, \ie when agents have a high enough communication demand.
%We observe that both the regular and extended PT result in lower~$\bar{E}$ than either ET design.
%It can also be seen that PT* is able to reduces the utilization, compared to the regular PT, while maintaining the control error.
%The PT framework also shows that resources remain available (\ie a network utilization well below 100\%) to either be shut off to save energy, or allocated for other purposes.
%\jm{Note to self: they assume the instantaneous receit of decisions, might be challenging if there are many agents}
%\vspace{-1mm}
\subsection{Stabilization}
\label{sec:Stabilization}
%A stabilization experiment is also considered, where all agents run a standard LQR controller with parameters $Q_{i}$, $R_{i}$ as above. Each agent run a standard LQR controller with parameters $Q_{i}$, $R_{i}$ as above.
Finally, we run a stabilization experimengt, where the feedback loop is closed over a network.
A heterogeneous system is considered, with simulated agent models as in \cite{WirelessFeedback}, and each agent runs~an~LQR controller with parameters $Q_{i}\mathrm{,}\,R_{i}$ as above.
An impulse disturbance is applied to the simulated agents at a random time $t_{d,i}\!\in\!\lbrack10,T\!-\!5\rbrack$ and to the physical agent at \SI{20}{\second}.
%The experiments demonstrate that stabilization of the physical agent is achievable for all triggering methods, down to $K=4$. When $K=3$, th
%The experiments demonstrate that a minimum of $K=2$ slots are required to stabilize the physical agent using the PT, whereas the ET1 design requires a minimum of $K=3$ slots.
The experiments demonstrate that $K\geq2$ slots are required to stabilize the physical agent with the PT, whereas $K\geq3$ slots are required for the ET1 design.
%The experiments demonstrate that in order to stabilize the physical agent, a minimum of $K=2$ slots are required for the PT, whereas the ET1 design requires $K=3$ slots.
For $K\!=\!2$, the physical agent goes unstable~prior to the disturbance, with ET1, but with the PT it remains stable throughout (see \url{https://sites.google.com/view/ptformas} for videos).
%\footnote{CACC simulation code (Matlab) for \secref{sec:CACC} and videos of the stabilization experiments (\secref{sec:Stabilization}), for $K\!=\!2$, are available at \url{https://sites.google.com/view/ptformas}\label{ft:PaperSite}}
%\footnote{CACC simulation code (Matlab) and videos of the stabilization experiments, for $K\!=\!2$, are available at \url{https://ics.is.mpg.de/employees/mastrangelo}\label{ft:MPIPage}}
%\db{we still need to figure out how we want to do this in the end, not sure what we decided on}
\vspace{-0.5em}
\section{Conclusion}
\label{sec:Conclusion}
\vspace{-0.5em}
This paper presents a resource constrained predictive triggering framework capable of allocating limited communication resources in multi-agent systems. 
By computing the exit probability of a stochastic process, a probabilistic measure is derived to prioritize resource allocation.
We have shown that by communicating this probability periodically, network utilization may be reduced while also improving the control performance of distributed systems.
%We have shown that by allocating resources in advance, the performance of distributed control systems with more agents than communication slots can be improved.
The effectiveness of the proposed framework is demonstrated in CACC platooning simulations and in experiments featuring real and simulated cart-pole systems.
%The effectiveness of the proposed framework is demonstrated in CACC platooning simulations and in experimental cart-pole systems.
%\vspace{-0.2em}
\begin{ack}
\vspace{-0.5em}
The authors would like to thank Joel Bessekon Akpo for his help with building the experimental platform and Friedrich Solowjow for helpful discussions.
\end{ack}

\bibliography{bibliography}             % bib file to produce the bibliography with bibtex (preferred)

\end{document}